\documentclass[journal=acscii,manuscript=article]{achemso}
\setkeys{acs}{usetitle = true}

\usepackage{amssymb,amsfonts,amsmath}

\author{Salvador Barraza-Lopez}
\email{sbarraza@uark.edu}
\affiliation{Department of Physics, University of Arkansas, Fayetteville, Arkansas 72701, United States}
\alsoaffiliation{Institute for Nanoscale Science and Engineering, University of Arkansas, Fayetteville, Arkansas 72701, United States}
\author{Thaneshwor P. Kaloni}
\affiliation{Department of Physics, University of Arkansas, Fayetteville, Arkansas 72701, United States}
\alsoaffiliation{Institute for Nanoscale Science and Engineering, University of Arkansas, Fayetteville, Arkansas 72701, United States}
\title{Water Splits to Degrade Two-dimensional Group-IV Monochalcogenides in Nanoseconds}

\begin{document}

\begin{tocentry}
\begin{center}
\includegraphics[width=1\textwidth]{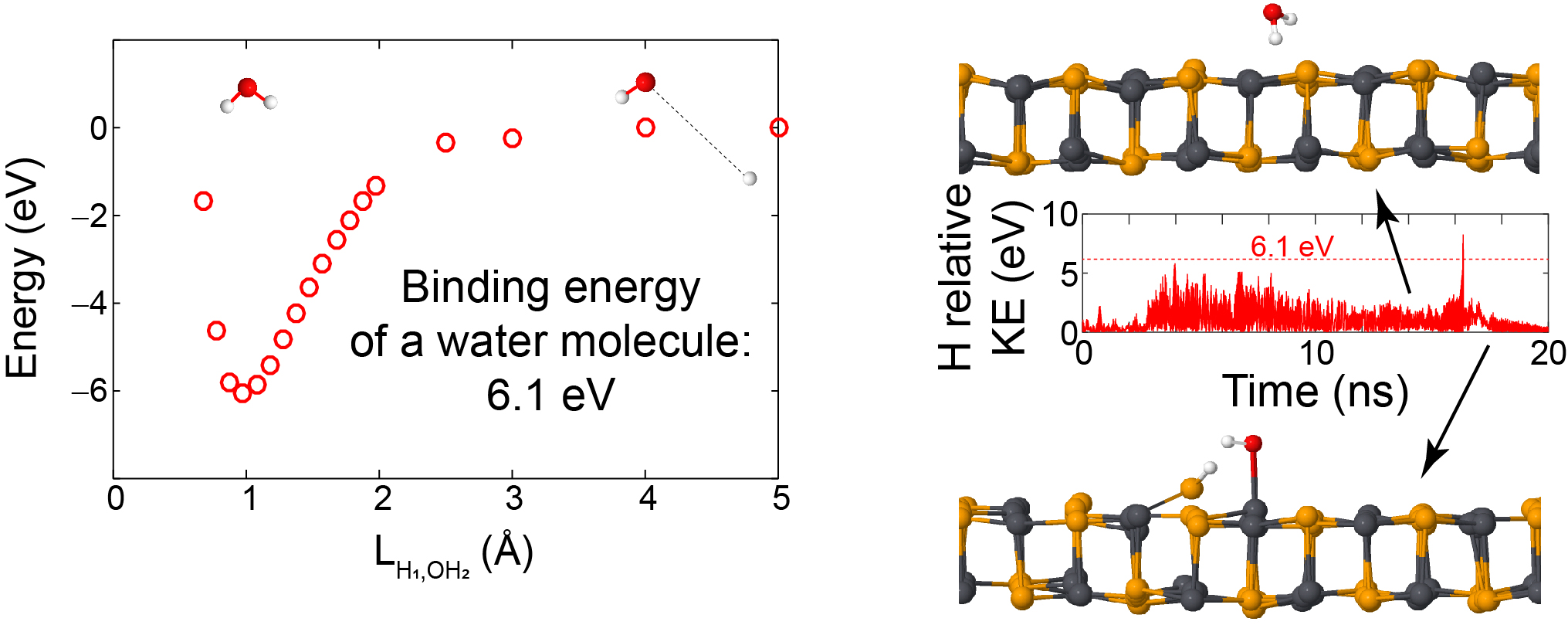}
\end{center}
Exfoliation of layered group-IV monochalcogenides on water-containing environments leads to their chemical decomposition within nanoseconds. These materials could be useful for hydrogen production.
\end{tocentry}

\begin{abstract}
The experimental exfoliation of layered group-IV monochalcogenides --semiconductors isostructural to black phosphorus-- using processes similar to those followed in the production of graphene or phosphorene has turned out unsuccessful thus far, as if the chemical degradation observed in black phosphorus was aggravated in these monochalcogenides. Here, we document a facile dissociation of water by these materials within ten nanoseconds from room-temperature Car-Parrinello molecular dynamics calculations under standard temperature and pressure conditions. These results suggest that humidity must be fully eradicated to exfoliate monolayers successfully, for instance, by placing samples in a hydrophobic solution during mechanical exfoliation. From another materials perspective, these two-dimensional materials that create individual hydrogen ions out of water without illumination may become relevant for applications in hydrogen production and storage. This document is the Accepted Manuscript version of Published Work that appeared in final form in ACS Central Science, copyright American Chemical Society after peer review and technical editing by the publisher. To access the final edited and published work please use the following DOI: 10.1021\/acscentsci.8b00589
\end{abstract}

\section{Introduction}

Water molecules have an intrinsic electric dipole of 1.84 Debye \cite{Sharpe} that can receive a significant kinetic feedback from materials with large spatial charge inhomogeneities such as polar binary compounds known as group-IV monochalcogenide monolayers. As each of the three atoms on a water molecule negotiate their best placement with a material that produces inhomogeneous electric fields and also moves about at finite temperature, they strongly accelerate/decelerate and gain/loose kinetic energy past their 6.1 electron-Volt (eV) (70,000 K) energy barrier, to produce an unexpected splitting of hydrogen bonds. The discovery --which will be discussed in what follows-- implies a facile chemical degradation, and may explain why it is apparently harder to preserve exfoliated samples of these materials than it is to maintain phosphorene in its original chemical and crystalline composition after exfoliation.

Group-IV monochalcogenides \cite{Littlewood1,Littlewood2,Kooi221} have demonstrated great potential for applications, commanding an interest from the Engineering, Materials, and Physics disciplines. Indeed, besides their uses in photovoltaic, \cite{photovoltaics1,antunez1,antunez2} photoelectric\cite{photodetector} and piezoelectric\cite{generator} --i.e., energy conversion\cite{energy0,energy00,energy3,energy1,wenjuan0}-- applications, these transducer materials have additional and unique qualities such as: a record-setting thermoelectric figure of merit \cite{Heremans554,Zhao2014,delaire,seebeck}, the possibility of realizing topological crystalline insulators (quantum materials in which electronic states are created at surfaces as a result of crystal symmetries) \cite{theoTCI,Tanaka2012}, and can also host in-plane ferroelectricity \cite{Mehboudi-nl,Chang274,prlferro,prbferro,Raad} in ultrathin layered samples, opening the door for a dedicated search of ferroic behavior in two-dimensional and layered materials \cite{Menghao-nl,2D-mater,new,new2,newest,Kai2018}. These materials realize shift-current photovoltaics and electronic valleys addressable with linearly-polarized light\cite{rodin,tanaka,Cook2017,PhysRevLett.119.067402,exptlinear}

In-solution fabrication of ultrathin nanoplates of GeS, GeSe, SnS and SnSe has been reported, \cite{GeS,chemmatGeSe4MLs,seifert,antunez2} Raman signatures of vibrational mode softening were provided for 4-10 monolayer-thick SnSe, \cite{ramansnse} and the first experimental demonstration of ferroic behavior and of two-dimensional structural phase transitions in monolayers of SnTe was performed on ultrathin samples grown in ultra-high vacuum.\cite{Chang274} The communities vested on unveiling the unique properties of ultrathin group-IV monochalcogenides will agree that progress on this field could be greatly improved by a fast and reliable route toward high-quality monolayers that remain chemically stable for hours.

But a fundamental problem lingers unaddressed: layered group-IV monochalcogenides are isostructural and isoelectronic to black phosphorus, which exfoliates down to monolayers.\cite{Peide,Koenig,Li2014,Castellanos} Yet, there are no reports of ultrathin layered group-IV monochalcogenides produced by mechanical exfoliation using techniques that proved successful for the exfoliation of black phosphorus. In these pages, we argue that a quick degradation may be behind the lack of reports on the exfoliation of group-IV monochalcogenides under ambient or glovebox conditions. On the other hand, these results may also represent a novel route to produce hydrogen out of water at room temperature and standard pressure, which could be of use for energy harvesting applications.

Black phosphorus becomes degraded by photo-oxidation at atomistic defects and at edges\cite{Barraza,uvstudy,Kaci2}, and it must be protected from ambient exposure\cite{hersam,alexandrareview}. As for group-IV monochalcogenides, there are three preceding theory works on the interaction of these materials with single atoms or molecules based on Density Functional Theory \cite{hk,ks,martin} calculations at zero temperature. The earliest work focuses on the effect of structural vacancies and of individual oxygen atoms on the electronic properties of GeS, GeSe, SnS and SnSe monolayers\cite{GomesOxidation}. There, oxygen dimers were split by hand onto individual oxygen atoms, prior to their placement in proximity of the 2D material.

In a second work, the reaction pathways for physisorbed oxygen dimers on pristine GeS, GeSe, SnS and SnSe monolayers were contrasted with the pathway on phosphorene, to conclude that none of these 2D materials splits O$_2$\cite{ZhouOxidation}. A third paper advocates for the breaking of water by SiS, SiSe, GeS, GeSe, SnS and SnSe monolayers under illumination. The interaction of water with these materials was captured indirectly, by means of free energies developed from energy barriers and electrostatic potentials obtained on individual unit cells.\cite{photocatalysis}

Chemical kinetics studies how matter changes --how some bonds weaken and new bonds form-- with time.\cite{Tro} Kinetics depend on three factors: (a) the concentration (density) of oxygen dimers or water molecules, (b) temperature and, crucial for the present purposes, (c) the {\em structure} and relative orientation of reacting bodies. Moving beyond interactions that formally occur at zero temperature \cite{GomesOxidation,ZhouOxidation,photocatalysis}, we study the time-dependent evolution of a single oxygen or water molecule in the gas phase around phosphorene and group-IV monochalcogenide monolayers under standard temperature and pressure in the absence of illumination.

Towards that goal, spin-polarized Car-Parrinello ({\em ab initio}) molecular dynamics\cite{car} calculations under the NPT ensemble (with a constant number of particles and target ambient pressure and room temperature) \cite{nose,hoover,pr1,pr2} were performed on 5$\times$5 periodic supercells, with a time resolution of 1.0 fs, for over 20 ns. These calculations include the thermal change of in-plane lattice constants that was described in previous work\cite{Mehboudi-nl,prlferro,prbferro} by construction, and are meant to represent monolayers of these materials that are freshly grown or exfoliated. A single oxygen dimer or water molecule is placed at a height midway a 20 \AA{}-high simulation box, and let to interact with two periodic images of the 2D material. Additional technical details can be found in the Methods Section. Considering the size of the simulation box, these single molecules build up a {\em concentration of roughly $10^{20}$/cm$^{3}$}, which is three orders of magnitude smaller than the density of liquid water, but two orders of magnitude larger than the water density at the dew point (details in Supplementary Information). In spite of all its approximations, the present study may reproduce experiment right after exfoliation in air and in dark conditions in the most faithful manner yet.

\section{Results}

\subsection{Dissociation energies of oxygen and water molecules}
\begin{figure}[tb]
\includegraphics[width=0.4\textwidth]{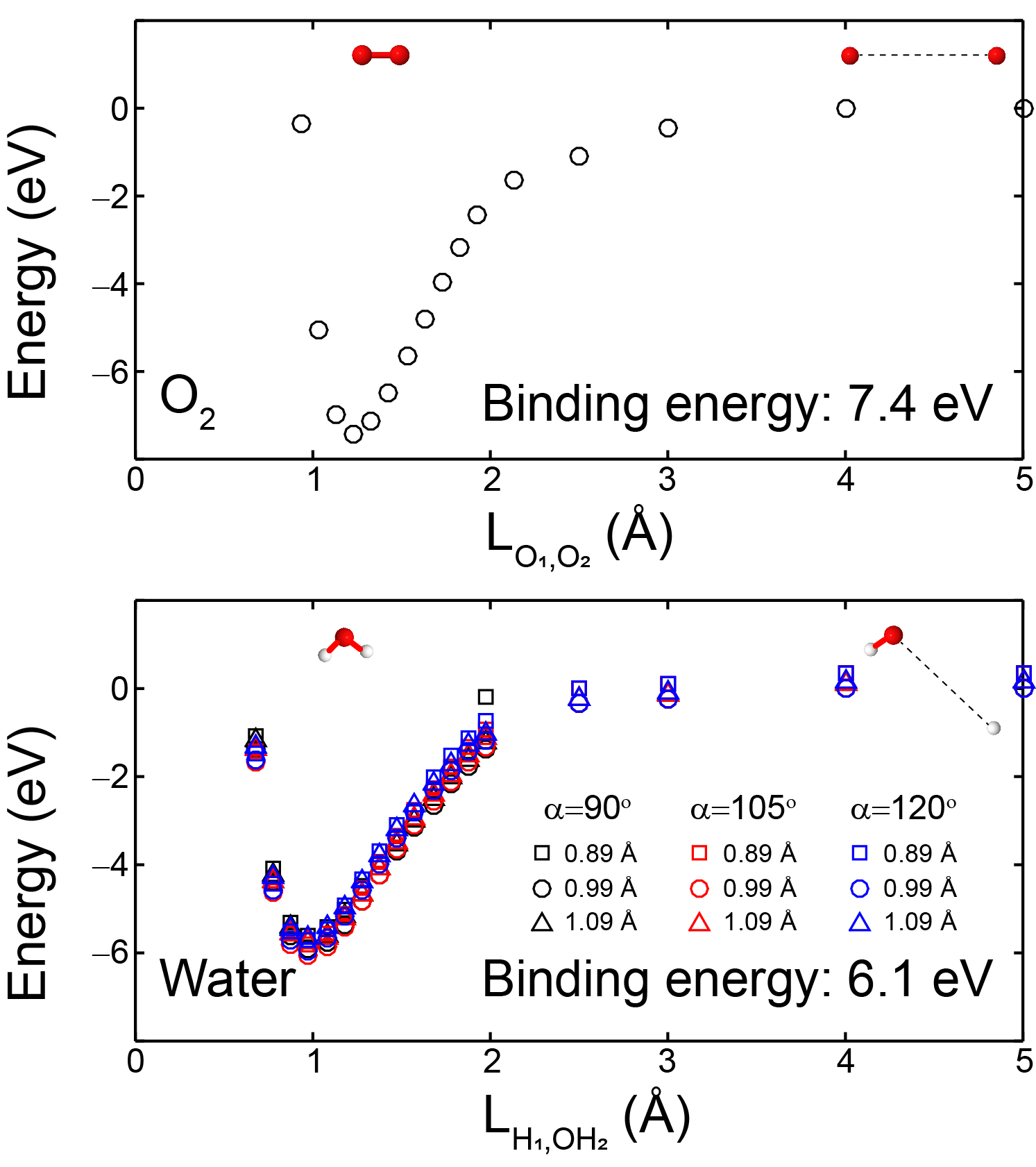}
\caption{Dissociation energy for oxygen and water molecules. For water, separation of a hydrogen atom was considered for three lengths of one O-H bond and for three values of the angle $\alpha$ formed by the H-O-H bonds, which was set when the two O-H bonds were equal.}
\label{fig:F1}
\end{figure}

Figure \ref{fig:F1} shows the dissociation energy of an oxygen dimer and a water molecule, including the effect of spin polarization. While the dissociation energy is only a function of distance for the dimer, $L_{O_1,O_2}$, dissociation of water means its splitting into an OH and a H fragment. And so, besides the distance from the separating hydrogen to the oxygen atom on the OH fragment (whose length $L_{H_1,OH_2}$ is kept fixed at will), we also considered variations on the length of the OH fragment ($L_{O,H_2}$) and on the angle $\alpha$, whose value was set when the two distances from hydrogens to the central oxygen atom were equal. Water dissociation is barely dependent on these two additional structural variables. The binding energies are 7.4 eV for the oxygen dimer, and 6.1 eV for water.

\subsection{Electrostatic potential of the 2D materials}

\begin{figure*}[tb]
\includegraphics[width=0.96\textwidth]{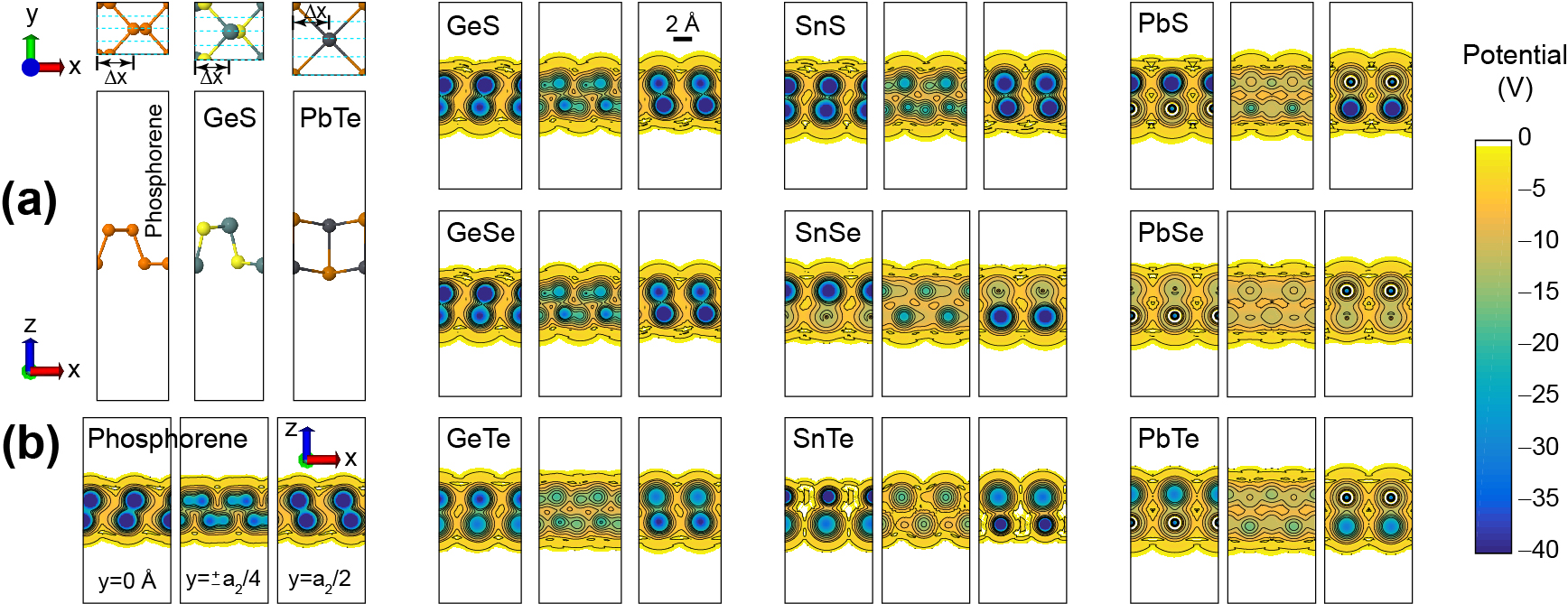}
\caption{(a) Atomistic structure of phosphorene, GeS and PbTe monolayers. (b) Side views of the electrostatic potential along the the cyan dashed lines in (a) emphasize the dissimilar electrostatic environment near phosphorene and group-IV monochalcogenide monolayers.}
\label{fig:F2}
\end{figure*}

Figure \ref{fig:F2} displays the electrostatic potential at the vicinity of the 2D materials at two-dimensional planes on the $x-z$ plane with the following fixed values of $y$: $0,\pm a_2/4$, and $a_2/2$ shown by dashed cyan lines on structural plots in Figure \ref{fig:F2}(a). (The larger degree of fluctuations on the electrostatic energy near Pb nuclei is due to the use of pseudopotentials with $d-$electrons promoted to the valence for that chemical element.)

In going from point ($0,0,z$) onto ($\Delta x,a_2/2,z$), with $\Delta x$ shown in Figure \ref{fig:F2}(a) as well, the electrostatic environment turns identical for (monoatomic) phosphorene. In IV-VI compounds, on the other hand, the electric field gradient aggravates along this cycle, because the nearest atoms at ($0,0,z$) and at ($\Delta x,a_2/2,z$) belong to different atomic species. This larger field gradient will prove crucial for the splitting of water.

\subsection{Kinetics of oxygen dimers and water molecules near the 2D materials}

Envision a black phosphorus or group-IV monochalcogenide monolayer just created, and place oxygen dimers or water molecules 8 \AA{} near the monolayer at room temperature and atmospheric pressure, to emulate the chemical interaction of these monolayers with proximal non-polar or polar molecules in the dark and under ambient conditions. The kinetics is uncovered by tracking energy-related and geometrical quantities such as the instantaneous pressure $P(t)$, temperature $T(t)$, configurational (DFT) energy $U(t)$, interatomic distances and angles throughout the molecular dynamics evolution.

\begin{figure*}[tb]
\includegraphics[width=0.96\textwidth]{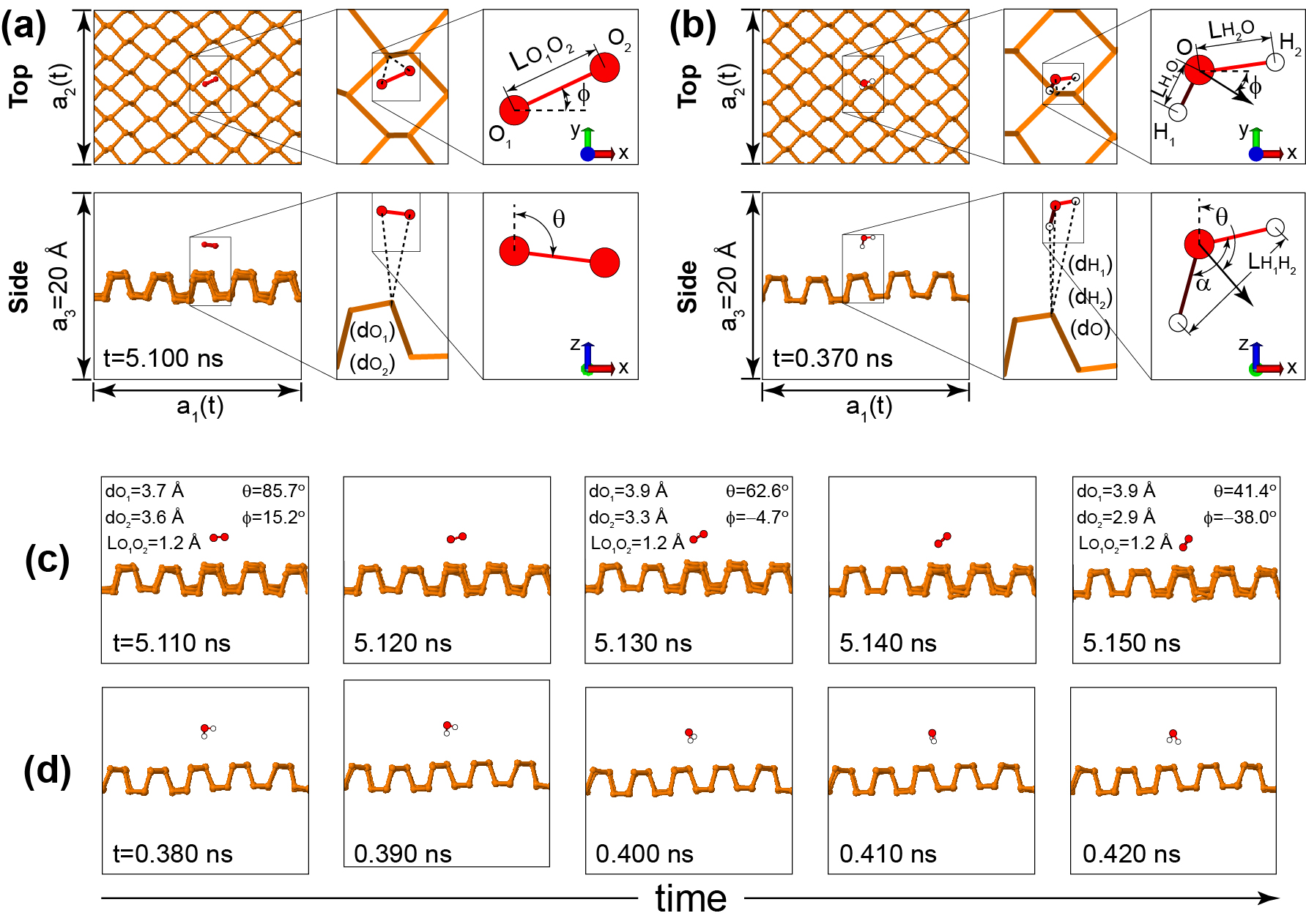}
\caption{Distances of (a) the oxygen dimer or (b) water to 2D materials (illustrated on phosphorene), and bond and angular distances to tell whether these two molecules split. The angle $\alpha $ and the orientation of water's dipole (angles $\theta$ and $\phi$) are included. Subplots (c) and (d) are still frames during the molecular dynamics evolution.}
\label{fig:F3}
\end{figure*}

The geometrical variables illustrated in Figures \ref{fig:F3}(a) and \ref{fig:F3}(b) are the smallest distance from individual oxygen and hydrogen atoms to the 2D material ($d_{O_1}(t)$ and $d_{O_2}(t)$ for the oxygen dimer; $d_{H_1}(t)$, $d_{H_2}(t)$ and $d_{O}(t)$ for water), the bond lengths of atoms of a given molecule ($L_{O_1,O_2}(t)$, or $L_{H_1,OH_2}(t)$), and the orientation of the water dipole given in terms of the polar angles ($\theta(t),\phi(t)$) defined such that the dipole orientation points from the negative to the positive net charge \cite{atkins}. Sudden, correlated changes in these kinetic quantities will highlight an undergoing chemical reaction, such as the breaking of molecular atomic bonds that would be signaled by magnitudes of $L_{O_1,O_2}(t)$ or $L_{H_1,OH_2}(t)$ much larger than those in a usual oxygen dimer or a water molecule, correlating with a temporary change in $T(t)$, and a permanent change on $U(t)$ post splitting.

Figures \ref{fig:F3}(c) and \ref{fig:F3}(d) are still frames from the phosphorene simulation that help understand the geometrical meaning of the structural variables within the molecular dynamics evolution. In Figure \ref{fig:F3}(c), a few frames on the evolution of the oxygen dimer are displayed; magnitudes of the geometrical variables displayed in three subplots assist in gaining intuition on the time evolution.

\subsubsection{Kinetics of oxygen dimers as a control study}

The oxygen dimer possess an intrinsic magnetic dipole moment of 2$\mu_B$, where $\mu_B$ is the Bohr magneton, that cannot interact with these non-magnetic 2D materials. The dimer will be employed to test whether the electric dipole of molecules interacting with these 2D materials plays a role on their splitting; it has a simpler structure than that of water --it is symmetric with respect to the vector joining the two oxygen atoms-- and due to its heavier mass it has a larger inertia as well.

\begin{figure*}[tb]
\includegraphics[width=0.96\textwidth]{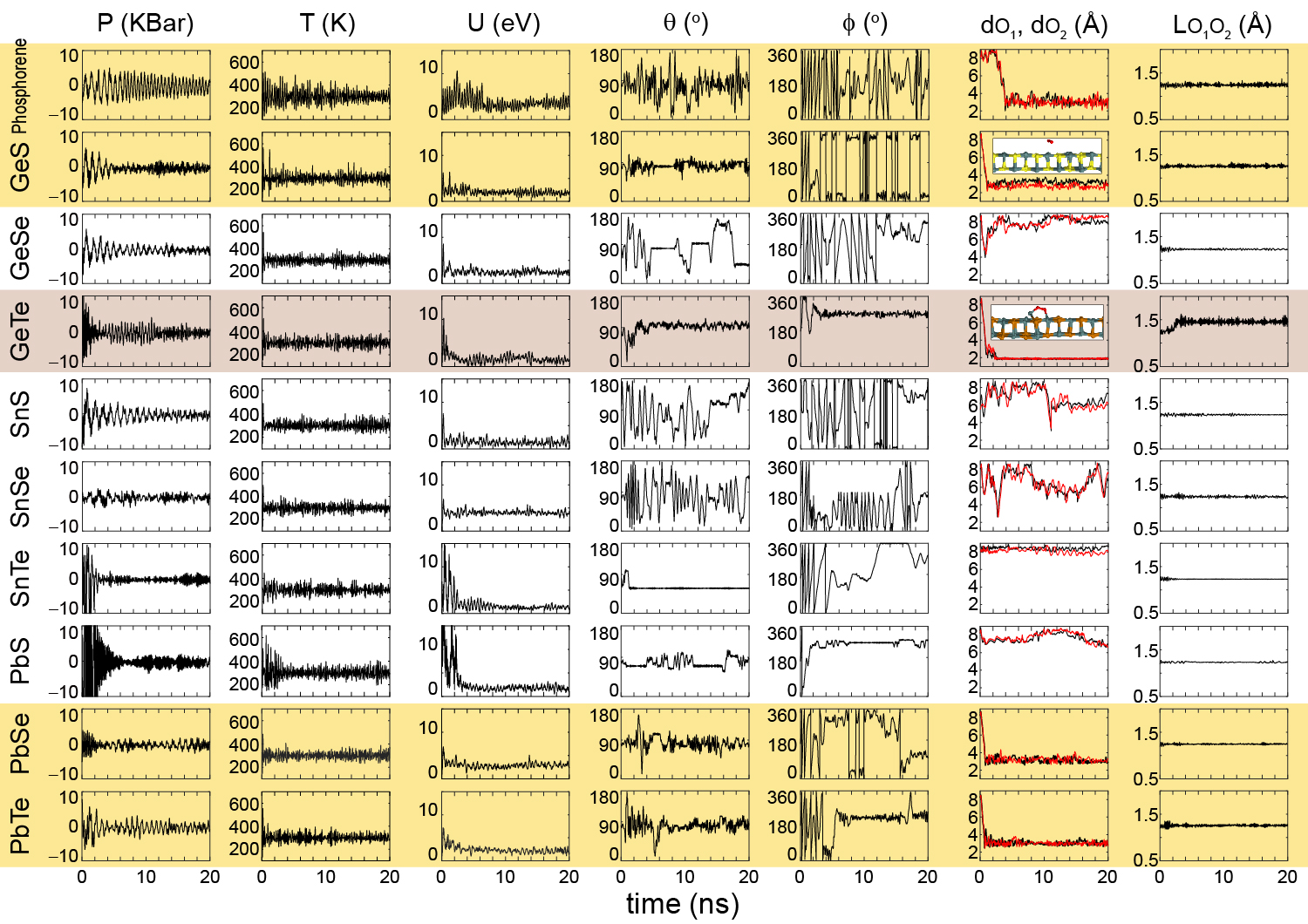}
\caption{Time evolution of an oxygen dimer in proximity of phosphorene and nine group-IV monochalcogenide monolayers. Lack of sudden and significant changes on $L_{O_1,O_2}$ indicate that defect-free samples do not break the oxygen bond within the simulation time. See main text for explanation of shading.}
\label{fig:F4}
\end{figure*}

Figure \ref{fig:F4} displays the magnitudes of $P(t)$, $T(t)$, $U(t)$, $\theta(t)$, $\phi(t)$, $d_{O_1}(t)$, $d_{O_2}(t)$, and $L_{O_1,O_2}(t)$ to understand the interaction among a dimer and phosphorene, and nine group-IV monochalcogenide monolayers. $P$, $T$ and $U$ stabilize as the simulation goes on, and provide trends to compare the evolution of water against later on. The angle $\theta$ is close to 90$^{\circ}$ for GeS, GeTe, SnTe, PbSe, and PbTe, implying a horizontal orientation. Nevertheless, $\theta$ shows larger fluctuations in phosphorene, GeSe, SnS, and SnSe, so the oxygen dimer rotates out of plane in the dynamical evolution involving these materials. The horizontal placement of the oxygen dimer  does not imply the formation of bonds to the 2D materials if $\phi$ is largely fluctuating too. Notice PbSe, where $\phi$ evolves largely over time, even when $\theta$ and the distances $d_{O_1}$ and $d_{O_2}$ to the 2D material are fixed. (Discontinuous jumps on $\phi$ imply an angular change between positive and negative angles around zero degrees.)

Subplots $L_{O_1,O_2}$ on the far right of Figure \ref{fig:F4} show no signs of breaking the oxygen bond up to the time the calculations stopped: the dimer maintains its length despite of its proximity to these 2D materials.  This observation can be further supported by the fact that spin remains equal to 2 in all of these calculations. As stated prior, oxidation processes will likely occur at structural defects, edges, and may even require optical excitations. One exception that we cannot explain, is the docking of the oxygen dimer onto GeTe determined by the reach of a constant height, lack of rotations (constant $\theta$ and $\phi$), and the increase of the dimer length past 1 ns. There, oxygen bonded to two germanium atoms, pulling a germanium above its original location in the 2D material, and its spin turns to zero, implying chemisorption (two characteristic atomistic frames are displayed in the $d_{O_1}$, $d_{O_2}$ subplots for GeS and GeTe).

\begin{table}
  \caption{Average distances from oxygen atoms to phosphorene and group-IV monochalcogenide monolayers, $d_{O_1}$  and $d_{O_2}$, and bond length $L_{O_1,O_2}$, collected from 4 to 20 ns.}
  \label{ta:t1}
  \begin{tabular}{c|cc|c}
    \hline
   Material & $d_{O_1}$  (\AA) & $d_{O_2}$ (\AA) & $L_{O_1,O_2}$ (\AA)\\
    \hline
Phosphorene  & 3.14$\pm$0.42 & 3.05$\pm$0.36 & 1.25$\pm$0.01\\

GeS          & 3.22$\pm$0.24 & 2.74$\pm$0.21 & 1.26$\pm$0.02\\
GeSe         & 8.06$\pm$0.41 & 8.13$\pm$0.49 & 1.23$\pm$0.01\\
GeTe         & 1.86$\pm$0.04 & 1.89$\pm$0.05 & 1.46$\pm$0.04\\

SnS          & 6.85$\pm$1.02 & 6.63$\pm$1.10 & 1.23$\pm$0.01\\
SnSe         & 6.64$\pm$1.15 & 6.59$\pm$1.04 & 1.23$\pm$0.02\\
SnTe         & 8.40$\pm$0.16 & 8.08$\pm$0.29 & 1.23$\pm$0.00\\

PbS          & 7.69$\pm$0.49 & 7.82$\pm$0.59 & 1.23$\pm$0.01\\
PbSe         & 3.13$\pm$0.25 & 3.20$\pm$0.25 & 1.25$\pm$0.01\\
PbTe         & 3.03$\pm$0.20 & 3.03$\pm$0.20 & 1.26$\pm$0.02\\

    \hline
  \end{tabular}

\end{table}

Table \ref{ta:t1} permits a classification of trends from the relative separation of oxygen atoms in the dimer to the 2D materials, despite of the identical initial placement, that is emphasized by the shading in Figure \ref{fig:F4}: GeSe, SnSe, SnSe, SnTe, and PbS are farthest away from the dimer and shown without shading. In fact, one sees in GeSe, SnS, and SnSe a sudden approach visible by sharp downward peaks that is aborted. Phosphorene, GeS, PbSe and PbTe have distances $d_{O_1}$ and $d_{O_2}$ oscillating in between 3.00 and 2.35 \AA, and emphasized with a yellow shading. Finally, GeTe, whose $d_{O_1}$ and $d_{O_2}$ are as short at 1.86 \AA{}, is shown with brown shading. None of these 2D materials in pristine form broke oxygen dimers and without the help from illumination within the simulation time.

\subsubsection{Water splitting}

Replacing the oxygen dimer by a water molecule, we contrast the interaction of a polar molecule and these materials against the behavior of the non-polar oxygen dimer. 

\begin{figure*}[tb]
\includegraphics[width=0.94\textwidth]{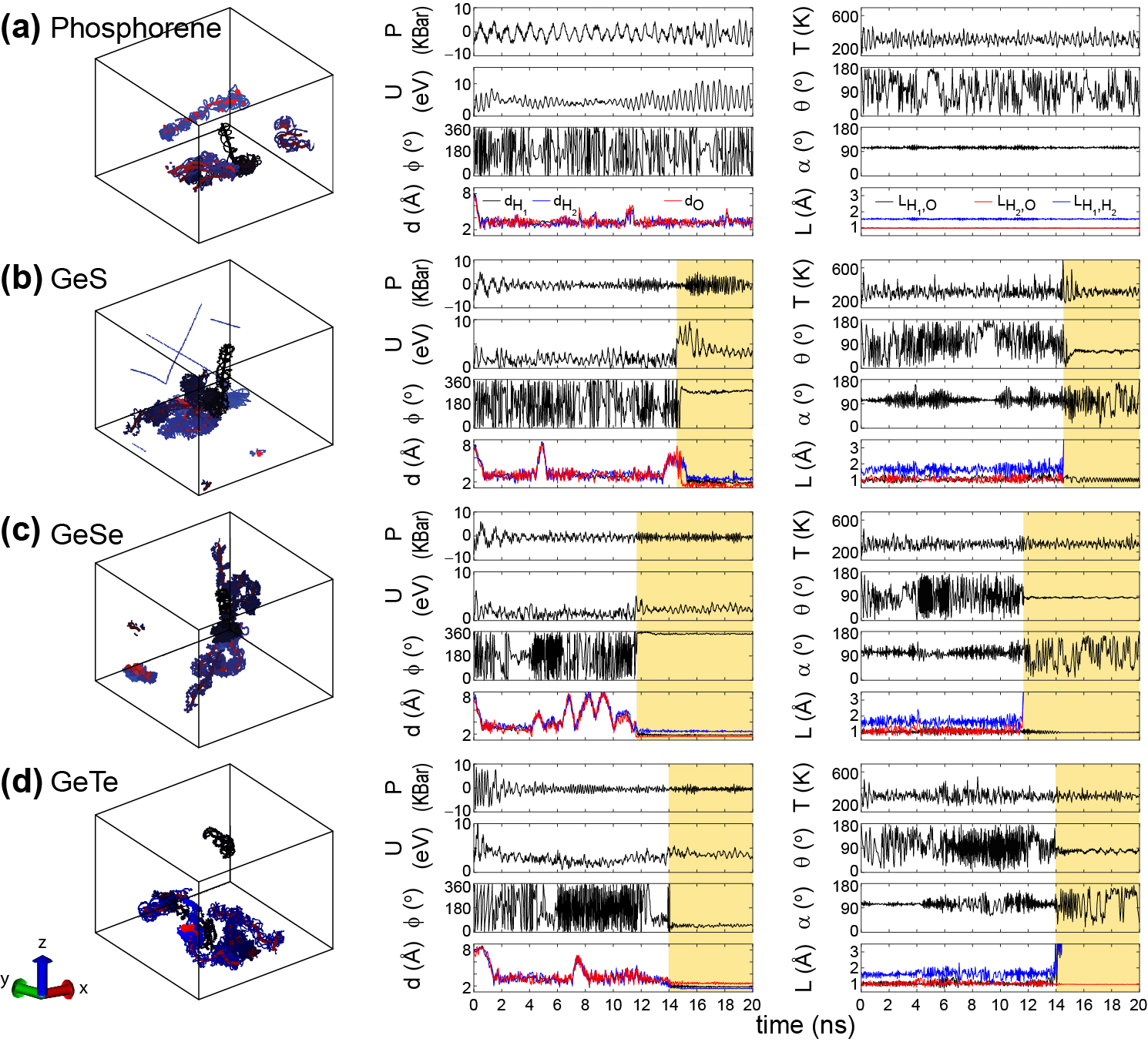}
\caption{Dynamics of (a) phosphorene and of (b) GeS, (c) GeSe, and (d) GeTe monolayers. Left: Trajectories of the three water atoms. Right: Evolution of energy-related and geometrical parameters. Ge-based monochalcogenide monolayers split water within 15 ns.}
\label{fig:F5}
\end{figure*}

In Figure \ref{fig:F5}(a) we show the evolution of a water molecule in proximity of (non-ferroelectric) phosphorene. Trajectories of the two hydrogen (blue) and the oxygen (red) atoms are shown at the leftmost side. Dark tones indicate earlier times, with black signaling $t=0$. Apparent discontinuities arise as the atoms move in between periodic images in the simulation box. The left panel in Figure \ref{fig:F5}(a) shows blue and red colors being tightly bound that imply a water molecule preserving its chemical integrity throughout the simulation, and the straight blue lines in the trajectory seen in Figure \ref{fig:F5}(b) provide the earliest indication of splitting.

\begin{figure*}[tb]
\includegraphics[width=0.94\textwidth]{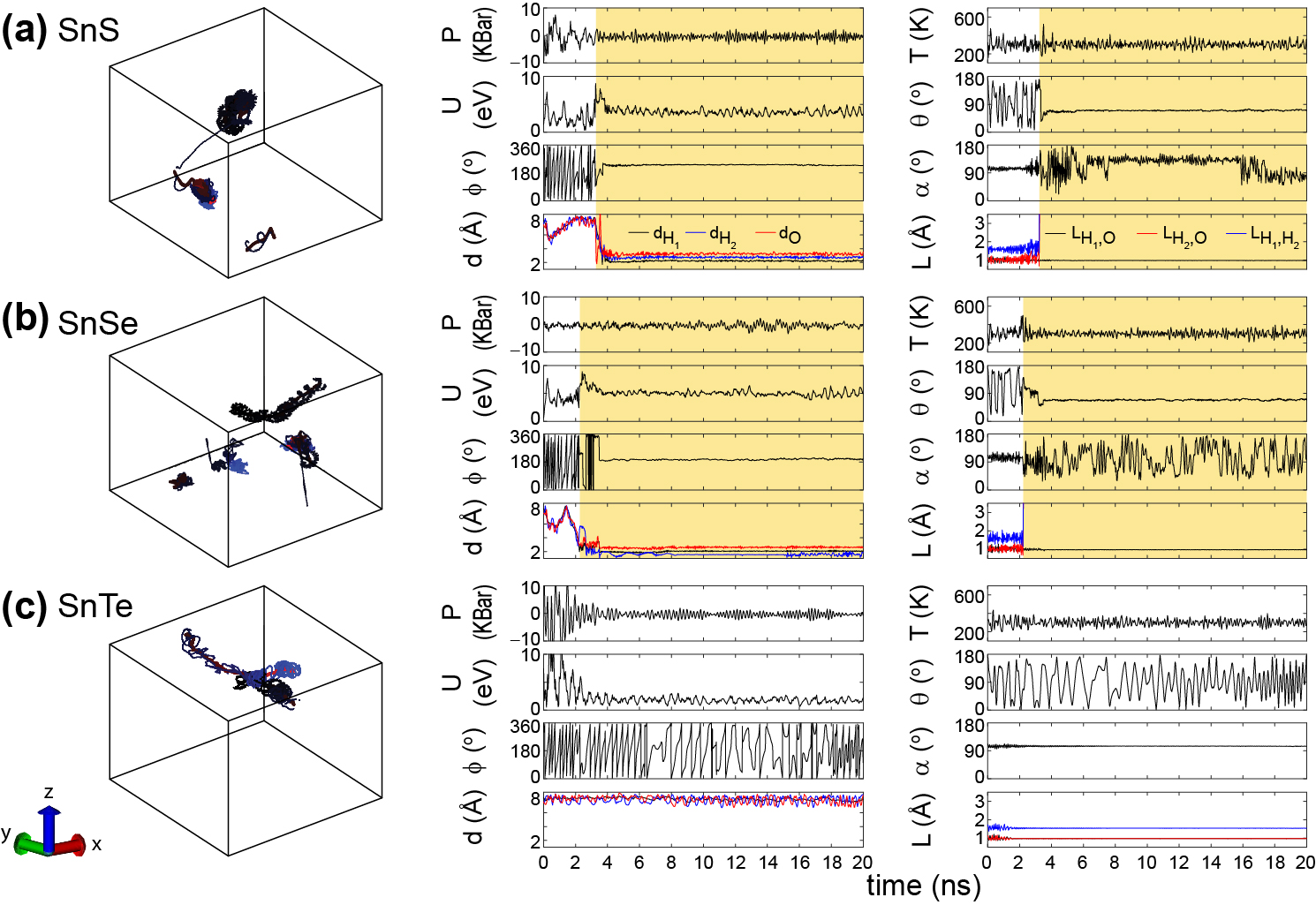}
\caption{Dynamics of (a) SnS, (b) SnSe, and (c) SnTe monolayers. Left: Trajectories of water atoms. Right: Evolution of energy-related and geometrical parameters. SnS and SnSe split water within 4 ns, while the heavier Sn-based compound did not break water.}
\label{fig:F6}
\end{figure*}

\begin{figure*}[tb]
\includegraphics[width=0.94\textwidth]{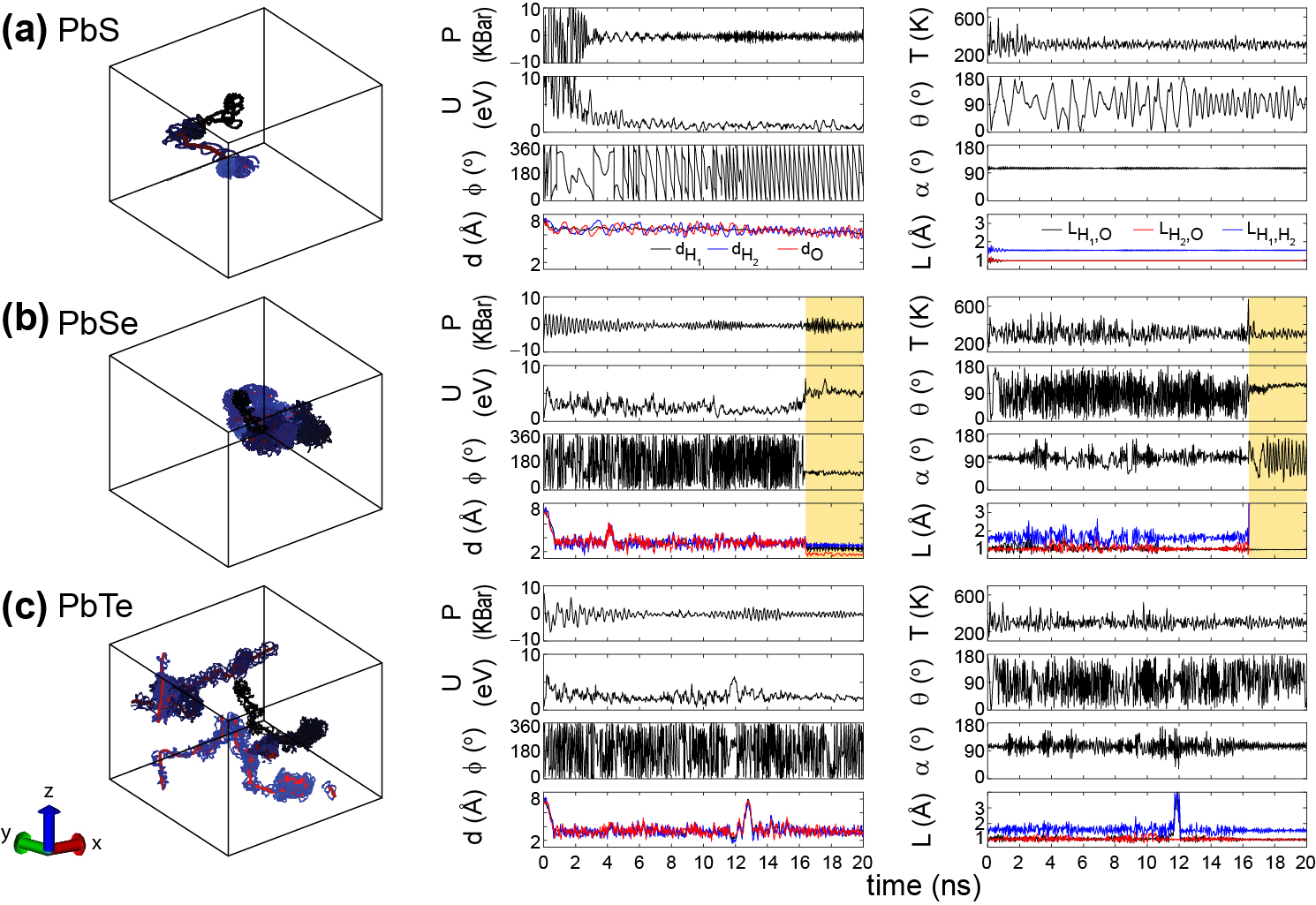}
\caption{Dynamics of (a) SnS, (b) SnSe, and (c) SnTe monolayers. Left: Trajectories of water atoms. Right: Parameter evolution. PbSe splits water past 16 ns.}
\label{fig:F7}
\end{figure*}

As in the case of the oxygen dimer (c.f. Figure \ref{fig:F4}), we tracked in Figures \ref{fig:F5} to \ref{fig:F7} $P(t)$, $T(t)$, $U(t)$, $\theta(t)$, $\phi(t)$, $\alpha$, $d_{O}(t)$, $d_{H_1}(t)$, $d_{H_2}(t)$, $L_{H_1,OH_2}$, and $L_{H_1,H_2}(t)$ as well. While $P(t)$ and $T(t)$ look similar to the control case shown in Figure \ref{fig:F4}, all other subplots can provide fingerprints of water splitting: sudden upward jumps on $U(t)$; lack of changes on $\theta$ and $\phi$ (which imply docking onto the 2D material); large fluctuations on $\alpha$ (as OH and H fragments find their placement in the 2D material) and on $L_{H_1,H_2}$. Shading emphasizes splitting.

Figure \ref{fig:F6} indicates that the two lightest Sn-based monochalcogenide monolayers, SnS and SnSe, split water too, with a time to split of about 4 ns that is even smaller than that observed in Figure \ref{fig:F5} for Ge-based monolayers. Moving onto compounds with larger atomic numbers, we observe water at mid-distance in between periodic images of the SnTe monolayer (see distances $d$ in Figure \ref{fig:F6}(c)). Lacking a larger interaction strength with the 2D material, water did not split there (see bond lengths $L$ in Figure \ref{fig:F6}(c) too).

As for Pb-based monolayers, and as seen in Figure \ref{fig:F7}(a), water stays mid-distance between the periodic images of PbS and it does not split. It splits when in proximity of PbSe (Figure \ref{fig:F7}(b)) at about 16 ns, and it splits to quickly re-bind at about 12 ns (see $L$ in Figure \ref{fig:F7}(c)) on PbTe. Figures \ref{fig:F5} to \ref{fig:F7} suggest that monochalcogenides with largest atomic numbers have a more difficult time splitting water within the timescale of this study.

\begin{table*}
  \caption{Distances from atoms to phosphorene and group-IV monochalcogenide monolayers $d_{H_1}$, $d_{H_2}$ and $d_{O}$, and structural variables $L_{H_1,O}$, $L_{H_2,O}$ and $\alpha$. When breaking of water occurs, the reported values are prior to splitting. Averages were collected in between 4 and 20 ns in cases when water did not break, and as indicated otherwise.}
  \label{ta:t2}
  \begin{tabular}{c|ccc|ccc}
    \hline
   Material  & $d_{H_1}$ (\AA) & $d_{H_2}$ (\AA) & $d_{O}$  (\AA)  & $L_{H_1,O}$ (\AA) & $L_{H_2,O}$ (\AA) & $\alpha$ ($^{\circ}$)\\
    \hline
Phosphorene
 & 3.21$\pm$0.59 & 3.31$\pm$0.65 & 3.28$\pm$0.46 & 0.98$\pm$0.01 & 0.98$\pm$0.01 & 105$\pm$4\\

GeS\textsuperscript{\emph{a}}
 & 3.51$\pm$1.08 & 3.44$\pm$1.12 & 3.40$\pm$1.02 & 1.05$\pm$0.16 & 1.10$\pm$0.20 & 106$\pm$20\\
GeSe\textsuperscript{\emph{a}}
 & 5.80$\pm$1.71 & 5.53$\pm$1.83 & 5.70$\pm$1.77 & 1.04$\pm$0.13 & 1.08$\pm$0.21 & 104$\pm$19\\
GeTe\textsuperscript{\emph{a}}
 & 3.34$\pm$0.71 & 3.35$\pm$0.67 & 3.24$\pm$0.55 & 1.05$\pm$0.13 & 1.03$\pm$0.10 & 105$\pm$22\\

SnS\textsuperscript{\emph{b}}
 & 7.86$\pm$0.63 & 7.77$\pm$0.60 & 7.99$\pm$0.58 & 0.99$\pm$0.06 & 1.06$\pm$0.17 & 104$\pm$11\\
SnSe\textsuperscript{\emph{b}}
 & 6.14$\pm$1.36 & 6.26$\pm$1.09 & 6.31$\pm$1.28 & 1.10$\pm$0.25 & 1.01$\pm$0.11 & 104$\pm$11\\
SnTe
 & 7.87$\pm$0.46 & 7.96$\pm$0.45 & 8.03$\pm$0.19 & 0.97$\pm$0.00 & 0.97$\pm$0.00 & 105$\pm$1\\

PbS
 & 6.77$\pm$0.54 & 6.73$\pm$0.49 & 6.75$\pm$0.23 & 0.97$\pm$0.00 & 0.97$\pm$0.00 & 105$\pm$2\\
PbSe\textsuperscript{\emph{a}}
 & 3.28$\pm$0.56 & 3.29$\pm$0.60 & 3.28$\pm$0.37 & 1.03$\pm$0.10 & 1.03$\pm$0.10 & 106$\pm$15\\
PbTe
 & 3.43$\pm$0.81 & 3.50$\pm$0.83 & 3.42$\pm$0.70 & 1.07$\pm$0.35 & 1.03$\pm$0.12 & 105$\pm$20\\
    \hline
  \end{tabular}

  \textsuperscript{\emph{a}} Split water molecule, averages over the 4,000 frames right before splitting time.\\
  \textsuperscript{\emph{b}} Split water molecule, averages over the 2,000 frames right before splitting time.
\end{table*}

Fluctuations of $d_{H_1}$, $d_{H_2}$ and $d_{O}$ in Table \ref{ta:t2} are smallest on materials that did not split water (phosphorene, SnTe, and PbS). The large magnitudes of $d_{H_1}$, $d_{H_2}$ and $d_{O}$ for SnS and SnSe are due to the ultra-fast splitting, that makes us include distances close to $t=0$ in the average. GeSe has the largest average distances to water and the largest fluctuations.

\section{Discussion}

The degradation of group-IV monochalcogenide monolayers is fundamentally different from that of phosphorene: the intrinsic in-plane electric dipole --responsible for the ferroelectric behavior of these materials-- attracts and splits water molecules even in the absence of structural defects or edges, and without the need for illumination, resulting on the chemical compromise and degradation within nanoseconds. The degradation hereby unveiled is more aggressive than that reported for black phosphorus, in which degradation requires the additional presence of structural defects or edges, and illumination. Our findings are in stark contrast with the conclusion that GeS, GeSe, SnS, and SnSe are chemically stable in an aqueous environment and under ambient electrochemical conditions.\cite{photocatalysis}

\begin{figure}[tb]
\includegraphics[width=0.48\textwidth]{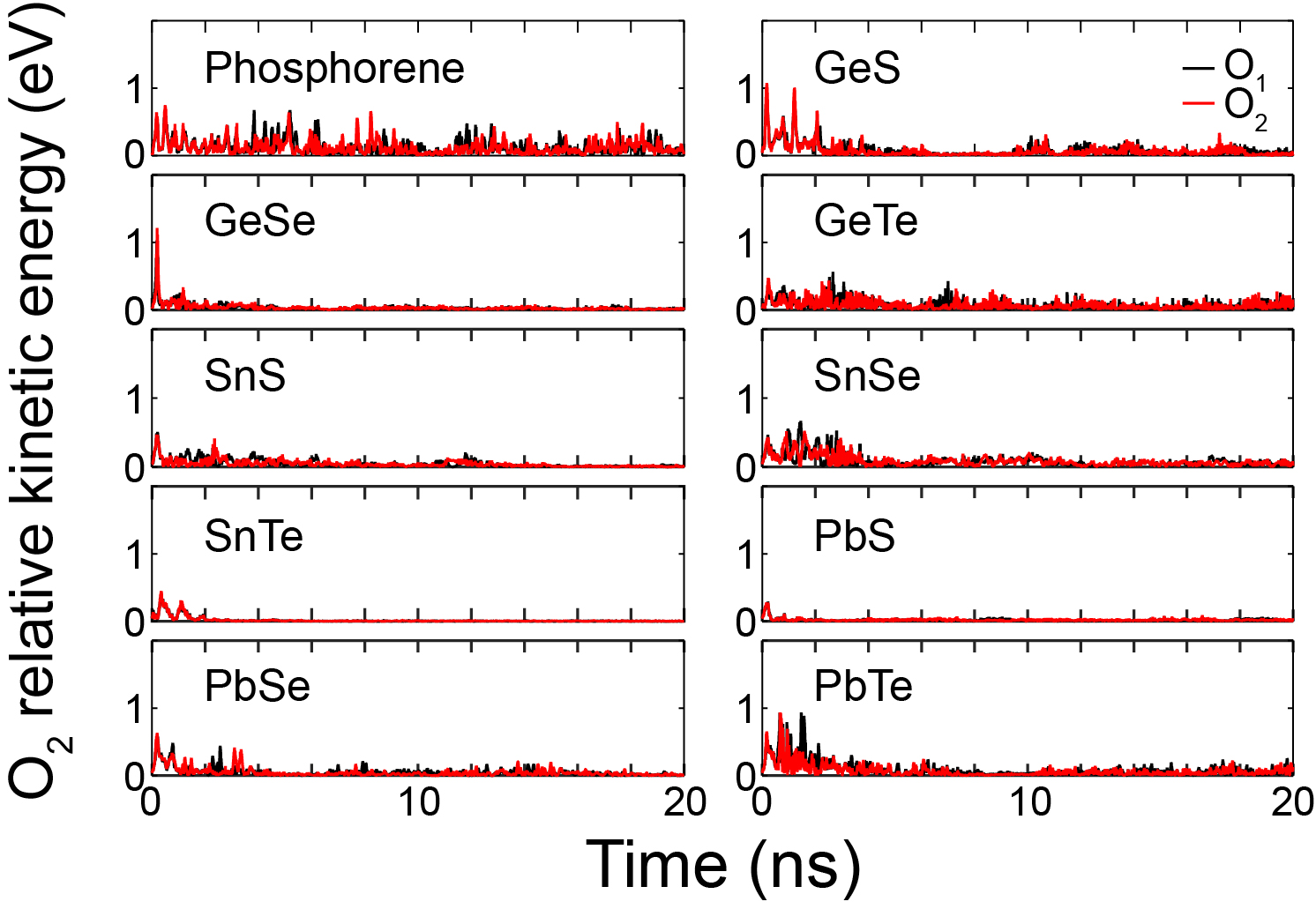}
\caption{Instantaneous kinetic energy ($KE$) for individual oxygen atoms, relative to the oxygen dimer's center of mass. The dimer would break if $KE> 7.4$ eV.}
\label{fig:F8}
\end{figure}

Phosphorene has five valence electrons and three chemical bonds, which creates lone pairs at every atom in its unit cell. These lone pairs are related to its sp$^3$ hybridization that gives rise to its layered conformation with a lower symmetry when compared to graphene, and to its higher chemical reactivity. In layered group-IV monochalcogenides, an electronic redistribution similar to a lone pair occurs on two out of the four atoms in the unit cell. \cite{Lefebvre,galy,stereo3,stereo4} Not being polar, pristine phosphorene appears unable to drive and accelerate polar molecules toward it (nor non-polar oxygen dimers for that matter) with such an ease.

\subsection{Hydrogen acquires enough kinetic energy to split off from water}

\begin{figure}[tb]
\includegraphics[width=0.48\textwidth]{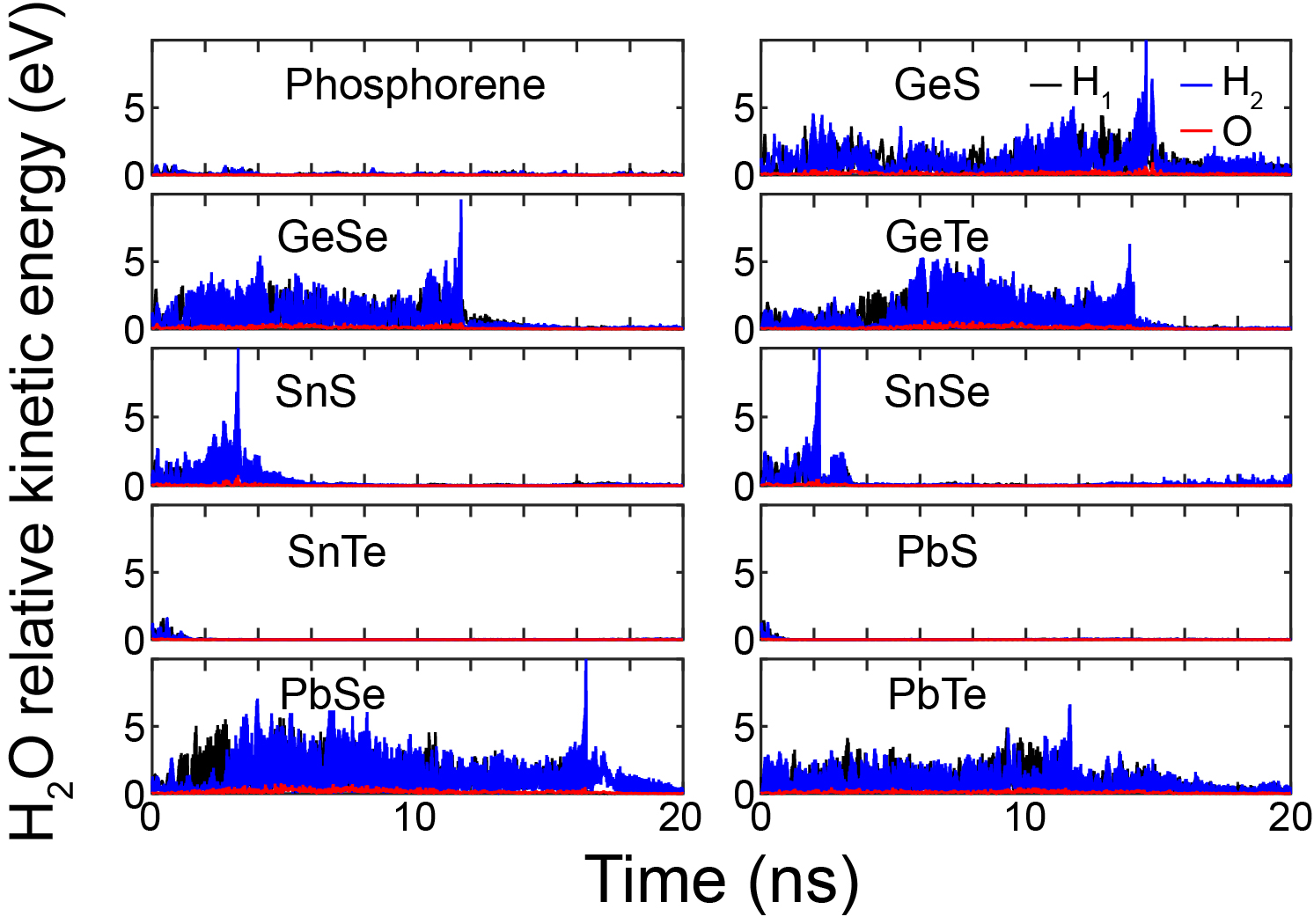}
\caption{Instantaneous kinetic energy ($KE$) for individual hydrogen and oxygen atoms, relative to water's center of mass kinetic energy. Water splits right when $KE>6.1$ eV.}
\label{fig:F9}
\end{figure}

Previous observation is established on a stronger footing in Figures \ref{fig:F8} and \ref{fig:F9}, via the kinetic energy of individual oxygen and hydrogen atoms belonging to the dimer or water molecules. We extract the kinetic energy relative to the center of mass motion to determine whether its magnitude is large enough to break molecular bonds (details are given in the Supporting Information). Figure \ref{fig:F8} indicates fluctuations of the kinetic energy smaller than 2 eV, that explain the lack of oxygen splitting in these runs.

The kinetic energy gained by water in Figure \ref{fig:F9}, a molecule with a net electric dipole, can become much higher in comparison when around these 2D materials. Phosphorene does not carry a net electric dipole, and the kinetic energy of water's hydrogen and oxygen atoms resembles the one observed for the oxygen dimer on this material. To drive this analog behavior home, in which a non-polar material interacts with polar and non-polar molecules similarly, the maximum kinetic energy relative to the center mass motion listed in Table \ref{ta:t3} has an almost identical magnitude for the dimer and water near phosphorene.

Fluctuations on the kinetic energy of hydrogen atoms on the water molecule in these NPT calculations can turn much larger than the thermal, mean kinetic energy (0.03 eV). Such an extreme build-up of the relative kinetic energy of the hydrogen atoms originates as the three atoms in the water molecule negotiate their best placement around the monochalcogenide monolayers. Unable to dock in these materials as a single molecular unit, they first break to create free radicals that dock soon afterwards, compromising the chemical stability of the 2D materials. Movies that highlight the interaction of water with these 2D materials for a tenth of the entire evolution are provided as Supplementary Information.

\begin{table}
  \caption{Maximum kinetic energy ($Max_{KE}$) for the oxygen dimer and water molecule with respect to the center of mass motion.}
  \label{ta:t3}
  \begin{tabular}{c|c|c}
    \hline
   Material & $Max_{KE}$, oxygen dimer (eV) & $Max_{KE}$, water (eV)\\
    \hline
Phosphorene & 0.82 & 0.86\\

GeS         & 1.05 & 14.92\\
GeSe        & 1.29 & 10.12\\
GeTe        & 0.56 &  6.32\\

SnS         & 0.65 & 16.98\\
SnSe        & 0.68 & 11.24\\
SnTe        & 0.48 &  1.74\\

PbS         & 0.33 &  1.91\\
PbSe        & 0.68 & 12.43\\
PbTe        & 1.03 &  6.64\\

    \hline
  \end{tabular}

\end{table}

\begin{figure}[tb]
\includegraphics[width=0.48\textwidth]{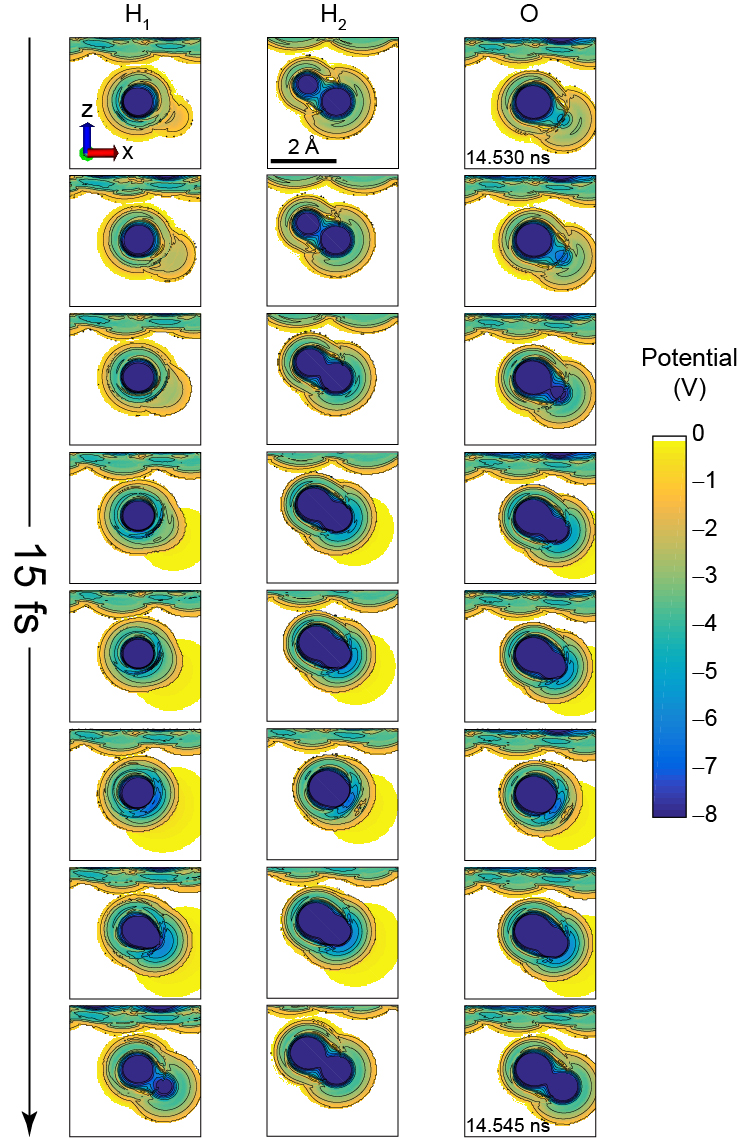}
\caption{The electrostatic energy centered about individual atoms on water during its interaction with GeS shortly before splitting.}
\label{fig:F10}
\end{figure}

Short of creating an effective model for the interaction of the dimer with the 2D materials, Figure \ref{fig:F10} shows the electrostatic energy of every atom in water shortly before the time at which water splits off GeS. To this end, we skipped every other frame to show 15 fs worth of dynamical evolution. The subplots are centered at the three individual atoms and contained within a 4$\times$4 \AA$^2${} simulation box. The observed fluctuations on the electrostatic energy around these atoms accelerate the atoms on the water molecule up to their splitting.

\subsection{Effect of water dissociation on electronic properties}

\begin{figure}[tb]
\includegraphics[width=0.48\textwidth]{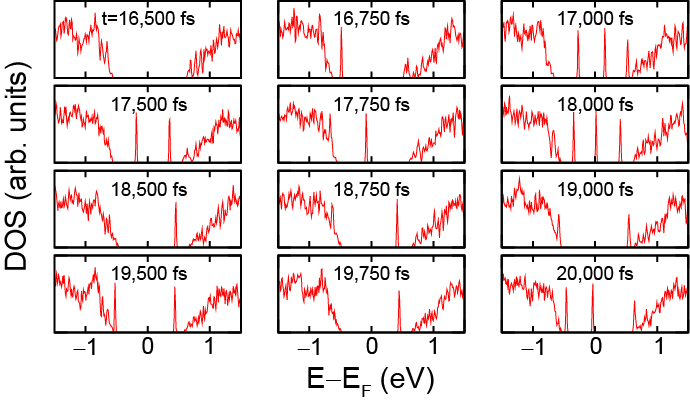}
\caption{Dynamical evolution of the electronic density of states for a water molecule after it splits and docks onto a PbSe monolayer.}
\label{fig:F11}
\end{figure}

\begin{figure*}[tb]
\includegraphics[width=0.96\textwidth]{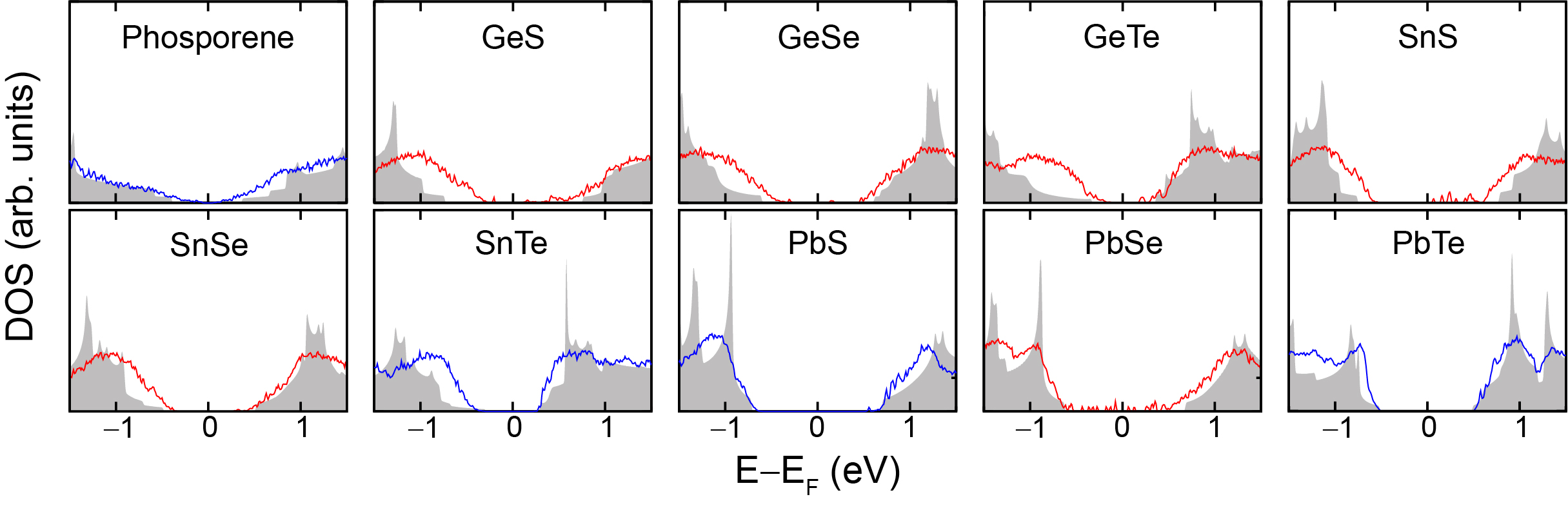}
\caption{Electronic density of states of the 2D materials with nearby water molecules. Area plots in gray show the density at zero temperature. The densities of states shown in red indicate the docking of water H and OH fragments onto the 2D material.}
\label{fig:F12}
\end{figure*}

Figure \ref{fig:F11} indicates a time dependency of the energy eigenvalue of the OH and H segments as they dock within the energy bandgap into the studied 2D materials, which could nowadays be verified by ultrafast probes. In Figure \ref{fig:F12}, the electronic density from such individual frames is added up, and shown for all the 2D materials we studied. The gray area plot in Figure \ref{fig:F12} is the reference density of states at zero temperature and without the water molecule. The gradual presence of states within the original bandgap indicates a degradation process that will continue as more and more water molecules are split and docked to the monochalcogenide monolayers that break water.

\begin{table}
  \caption{Some polar molecules, with their dipole in Debyes.}
  \label{tbl:t4}
  \begin{tabular}{c|c}
    \hline
   Molecule & Dipole (Debye)\\
    \hline
Water & 1.84\\
HF    & 1.82\\
HCl   & 1.08\\
HBr   & 0.44\\
CH$_3$Cl & 8.54\\
    \hline
  \end{tabular}
\end{table}

\subsection{Possibility to protect group-IV monochalcogenide monolayers from degrading}

We speculate that the facile degradation discovered here is behind the difficulty in exfoliating monolayers of group-IV monochalcogenides, and suggest employing non-polar and hydrophobic solvents removable by sonication to overcome the chemistry hereby demonstrated. This may help provide a rationale for the synthesis of these materials from --hydrophobic-- solutions \cite{chemmatGeSe4MLs,antunez2}. Non-polar capping coatings, when applied immediately after exfoliation, could also help preserve these ultrathin materials.

We also propose that other simple polar molecules, such as those listed in Table \ref{ta:t3} may behave similar to water molecules, and thus degrade these materials too. Working with specific --polar and non-polar-- molecular environments, and with the help of ultrafast probes, the findings here provided could be verified experimentally. In doing so, it may be possible to develop processes to protect these fascinating materials immediately after exfoliation, or to use them to produce hydrogen under gentle thermal and pressure conditions.

\section{Conclusion}

We discovered a dissociation of water molecules on group-IV monochalcogenide monolayers driven by abrupt local changes on the potential energy that the three constituent atoms of moving polar water molecules negotiate with. The steadfast breaking of water may be behind the difficulty to create ultrathin monochalcogenides from mechanical exfoliation that remain stable in between exfoliation and characterization. Conversely, the detailed study of the interaction of these materials with water may uncover a new energy material platform for the direct capture of hydrogen out of water in atom-thick membranes.

\section{Methods}\label{methods}
Spin-polarized Car-Parrinello molecular dynamics (MD) calculations were performed at 300 K on  NPT ensemble employing {\em SIESTA} package\cite{siesta}. Calculations were performed on 5$\times5$ supercells up to 20,000 \textit{fs}, with 1 \textit{fs} time resolution. Standard, DZP basis sets\cite{DZP}, and non-conserving Troullier-Martins\cite{TM} pseudopotentials with van der Waals corrections due to Berland and Per Hyldgaard\cite{PhysRevB.90.075148} and implemented by Rom\'an and Soler\cite{RS} that were tuned in-house\cite{rivero} were employed. All molecular dynamics calculations were performed with an identical Nos\'e Mass of 1500.0 Ry$\times$fs$^2$, and a Parrinello Rahman Mass of 1500.0 Ry$\times$fs$^2$. The lower and upper walls are kept fixed in the NPT evolution by an in-house modification to the \texttt{dynamics.f} routine. A sampling of $2\times2\times1$ $k-$points was used, with a precision in the electronic density of 10$^{-3}$. A real space grid with a 150 Ry cutoff was employed.

\section{Safety Statement}
No unexpected safety hazards were associated with the reported work.

\section{Associated content}
\noindent{\bf Supporting Information}\\
The Supporting Information is available free of charge on the ACS Publications website at DOI:\\
A discussion of the water density in our simulations, a zoom in of the variables shown in Figure \ref{fig:F5} for GeS near the time of splitting, and a description of movies showing the interaction of oxygen dimers and water molecules with black phosphorene and group-IV monochalcogenide monolayers (PDF).\\
Movies S1 to S10 corresponding to Figures 5 to 7.\\

\section{Author information}
\noindent{\bf Corresponding author}\\
*E-mail: sbarraza@uark.edu\\

T.P.K. carried out molecular dynamics, computed the potential energy, the spin-polarized DOS, and created the molecular dynamics movies. S.B.-L. carried out the analysis of the molecular dynamics data, crafted the figures, and wrote the manuscript.

\noindent{\bf Notes}\\
The authors declare no competing financial interest.

\section{Acknowledgments}
Research supported by the U.S. Department of Energy, Office of Basic Energy Sciences, under award  DE-SC0016139. Calculations were performed at Cori (NERSC), Carbon (ANL), and Trestles (Arkansas). Conversations with Kai Chang are gratefully acknowledged.



%



\providecommand{\latin}[1]{#1}
\providecommand*\mcitethebibliography{\thebibliography}
\csname @ifundefined\endcsname{endmcitethebibliography}
  {\let\endmcitethebibliography\endthebibliography}{}

\end{document}